\documentclass[aps, twocolumn, pre, superscriptaddress]{revtex4-1}
\usepackage{graphicx}
\usepackage{psfrag}
\usepackage{color}
\usepackage{latexsym}

\def\d {{\rm d}}

\begin{document}
\title{Macroscopic Discontinuous Shear Thickening vs Local Shear Jamming in Cornstarch}
\author{A. Fall}
\author{F. Bertrand}
\author{D. Hautemayou}
\author{C. Mezi\`ere}
\author{P. Moucheront}
\author{A. Lema\^{\i}tre}
\affiliation{Laboratoire Navier (UMR CNRS 8205), Universit\'e Paris Est, Champs-sur-Marne, France}
\author{G. Ovarlez}
\affiliation{Laboratoire Navier (UMR CNRS 8205), Universit\'e Paris Est, Champs-sur-Marne, France}
\affiliation{University of Bordeaux, CNRS, Solvay, LOF, UMR 5258, 33608 Pessac, France}

\date{\today}

\begin{abstract}
We study the emergence of discontinuous shear-thickening (DST) in cornstarch, by combining macroscopic rheometry with local Magnetic Resonance Imaging (MRI) measurements. We bring evidence that macroscopic DST is observed only when the flow separates into a low-density flowing and a high-density jammed region. In the shear-thickened steady state, the local rheology in the flowing region, is not DST but, strikingly, is often shear-thinning. Our data thus show that the stress jump measured during DST, in cornstach, does not capture a secondary, high-viscosity branch of the local steady rheology, but results from the existence of a shear jamming limit at volume fractions quite significantly below random close packing.
\end{abstract}

\maketitle

 Granular suspensions  are seemingly simple systems composed of non-colloidal rigid particles immersed in Newtonian fluids. Although the only relevant interactions are hydrodynamic and contact forces, they present a rich rheology~\cite{DennMorris2014}, which includes shear-thinning or shear thickening, normal stress differences, shear banding and yield stress behaviour. With increasing packing fractions, continuous shear thickening (CST) appears at lower and lower strain rates and becomes more and more abrupt. This eventually leads to the emergence of a spectacular phenomenon: discontinuous shear thickening (DST), an order-of-magnitude jump of the stationary (macroscopic) stress when strain rate crosses some threshold value. DST is observed in a narrow range of packing fractions near random close packing and recent works suggest that it is a dynamic or shear jamming transition. This motivates a considerable attention from experiments~\cite{BrownJaeger2009,NagahiroNakanishiMitarai2013,FernandezManiRinaldiKadauMosquetLombois-BurgerCayer-BarriozHerrmannSpencerIsa2013}, numerical simulation~\cite{SetoMariMorrisDenn2013,Heussinger2013,MariSetoMorrisDenn2014} and theory~\cite{NakanishiNagahiroMitarai2012,WyartCates2014}.

The origin of shear thickening is essentially unknown. Proposed mechanisms include: the crossover to Bagnold scaling due to particle inertia~\cite{FallLemaitreBertrandBonnOvarlez2010}; the coupling between normal and shear stresses associated with material dilatancy~\cite{FallHuangBertrandOvarlezBonn2008,BrownJaeger2012}. In numerical simulations DST was observed in systems which are forced to remain homogeneous by periodic boundary conditions~\cite{SetoMariMorrisDenn2013}. It was thus suggested that it might arise from a jump of the local response, which occurs at fixed packing fraction, and is caused by the proliferation of frictional contacts~\cite{SetoMariMorrisDenn2013,WyartCates2014}.

Examples abound, however, in granular suspension rheology where the macroscopic response is sharply different from the local behaviour: the apparent yield stress~\cite{OvarlezBertrandRodts2006,FallBertrandOvarlezBonn2009} or a transient DST behavior~\cite{FallLemaitreBertrandBonnOvarlez2010}, for example, were both shown to result from the emergence of flow inhomogeneities due respectively to density differences and migration. The observation of finite size effects in DST measurements~\cite{FallHuangBertrandOvarlezBonn2008,FallBertrandOvarlezBonn2012,BrownJaeger2012}, at sizes up to 100 particles, thus questions the idea that DST is a purely local phenomenon. In this context, it is unclear how a phase diagram for the local response can be deduced from macroscopic experimental data and compared with numerical~\cite{SetoMariMorrisDenn2013}, and theoretical~\cite{WyartCates2014} works.

This situation motivates us to study DST in cornstarch~\cite{MerktDeeganGoldmanRerichaSwinney2004,WagnerBrady2009,WaitukaitisJaeger2012}, the paradigmatic system for this phenomenon, by accessing experimentally the local response. To this aim, we have designed a new velocity-controlled rheometer that can exert steady torques up to 10 N.m on a wide-gap Couette cell inserted in our Magnetic Resonance Imaging (MRI) scanner. It enables us to access local velocity and particle volume fraction profiles in the flowing sample~\cite{OvarlezBertrandRodts2006,BonnRodtsGroeninkRafaiShahidzadeh-BonnCoussot2008}, for a wide range of strain rates ($\dot\gamma$) covering CST and DST. We thus find that macroscopic DST is observed when the flow localizes, i.e. the system separates into a low-density flowing and a high-density jammed region. In the shear-thickened steady state, the local rheology in the flowing region is not DST but, strikingly, is most often shear-thinning. Our data support that DST should be attributed to the existence of a shear jamming limit $\dot\gamma_{\rm DST}(\phi)$ at volume fractions $\phi$ significantly smaller than random close packing. 

The cornstarch (Sigma) is the same as in~\cite{MerktDeeganGoldmanRerichaSwinney2004,FallHuangBertrandOvarlezBonn2008}: it presents slightly polydisperse particles of size $\sim20\mu$m with irregular shapes. We have measured the random close packing fraction $\phi_{\rm RCP}\simeq55\%$ by tapping the powder for long time in a container~\cite{PhilippeBideau2002}. We have also measured the random loose packing $\phi_{\rm RLP}\simeq35\%$ as the density achieved after sedimentation of a dilute suspension~\cite{DongYangZouYu2006,CiamarraConiglio2008}. The suspensions are prepared by mixing cornstarch in demineralized water at initial volume fractions ranging from 33.5\% to 43.9\%.

\begin{figure}[t]
\includegraphics*[width=0.23\textwidth]{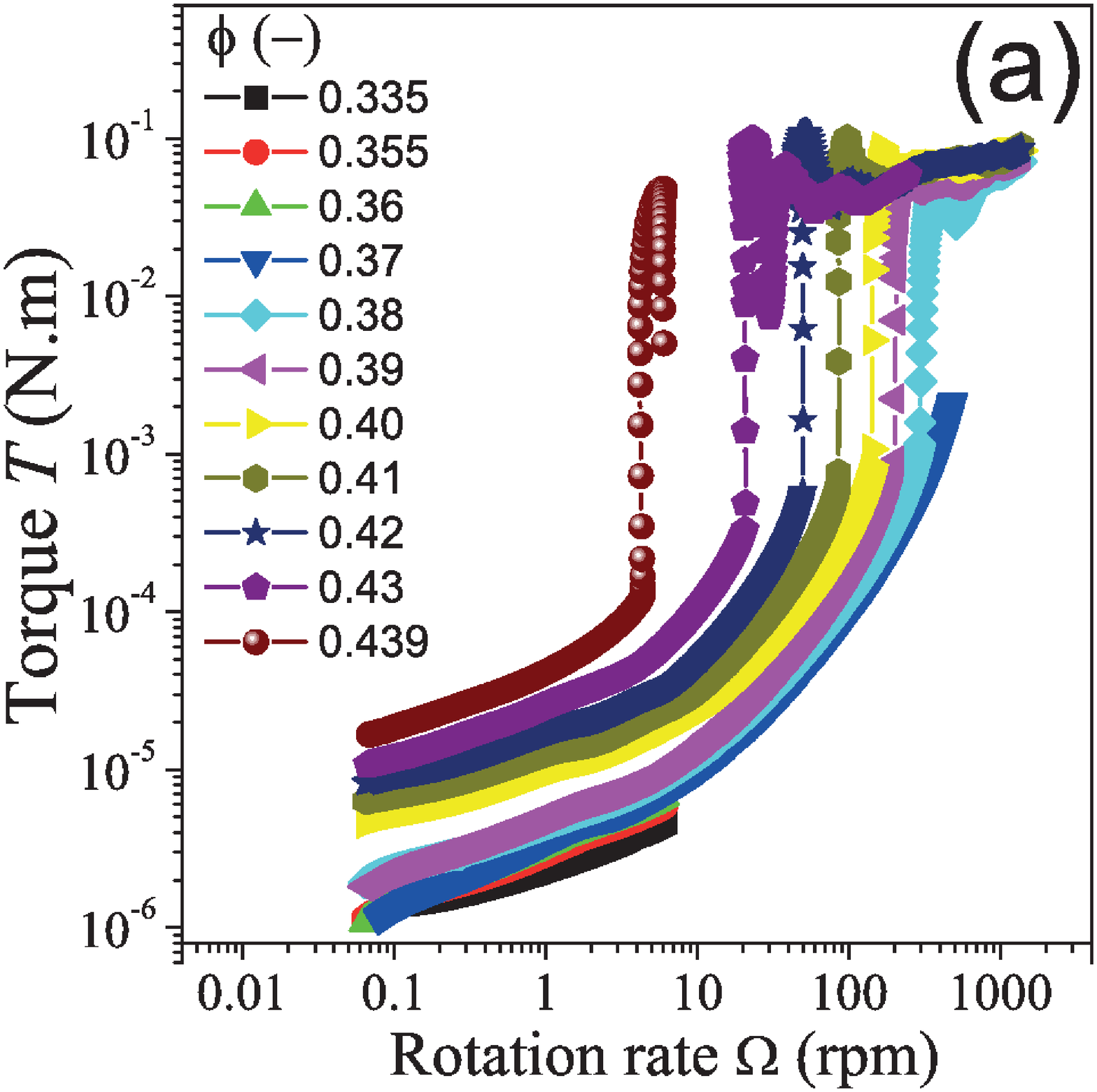}\hfil
\includegraphics*[width=0.23\textwidth]{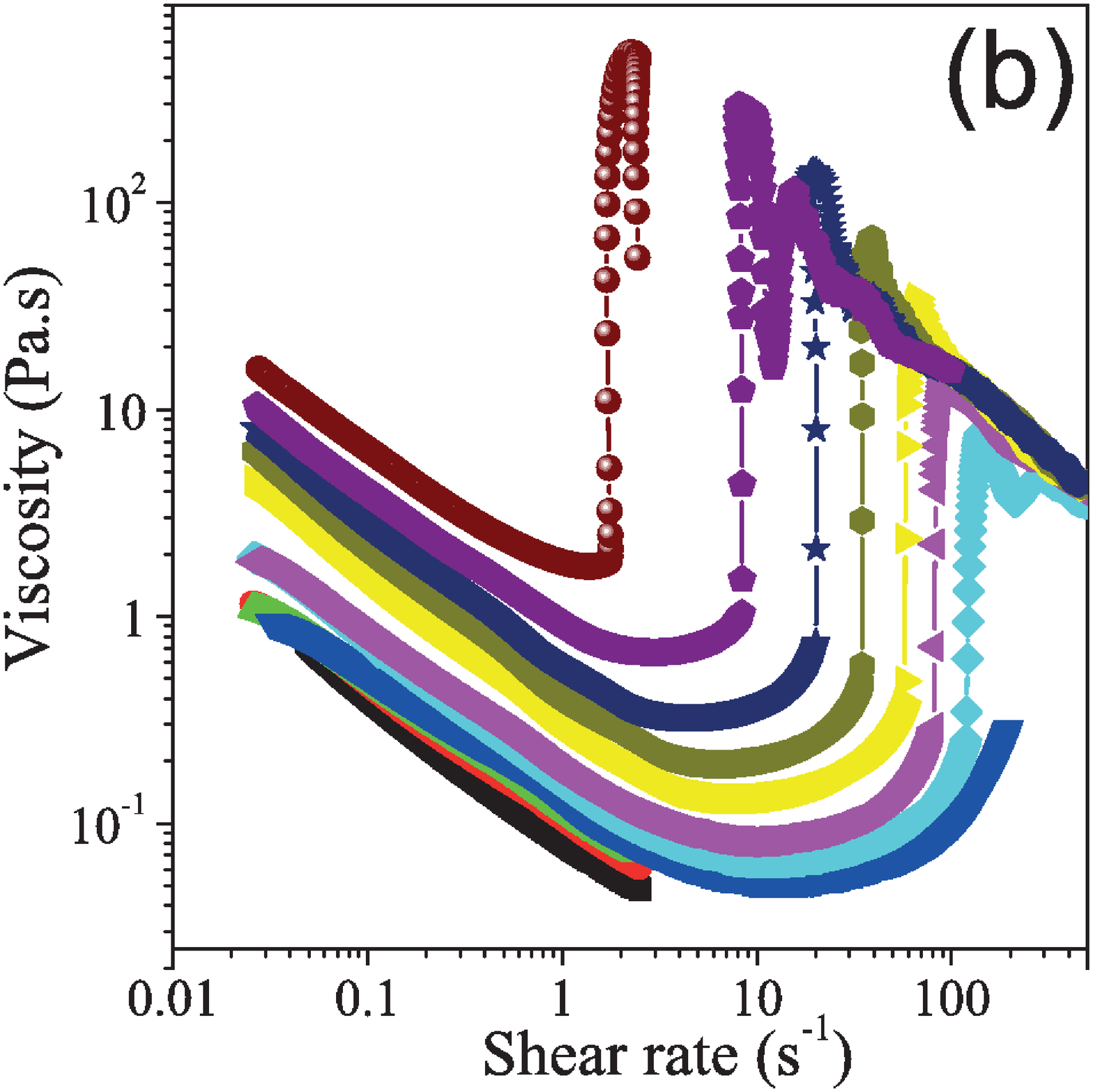}
\includegraphics*[width=0.23\textwidth]{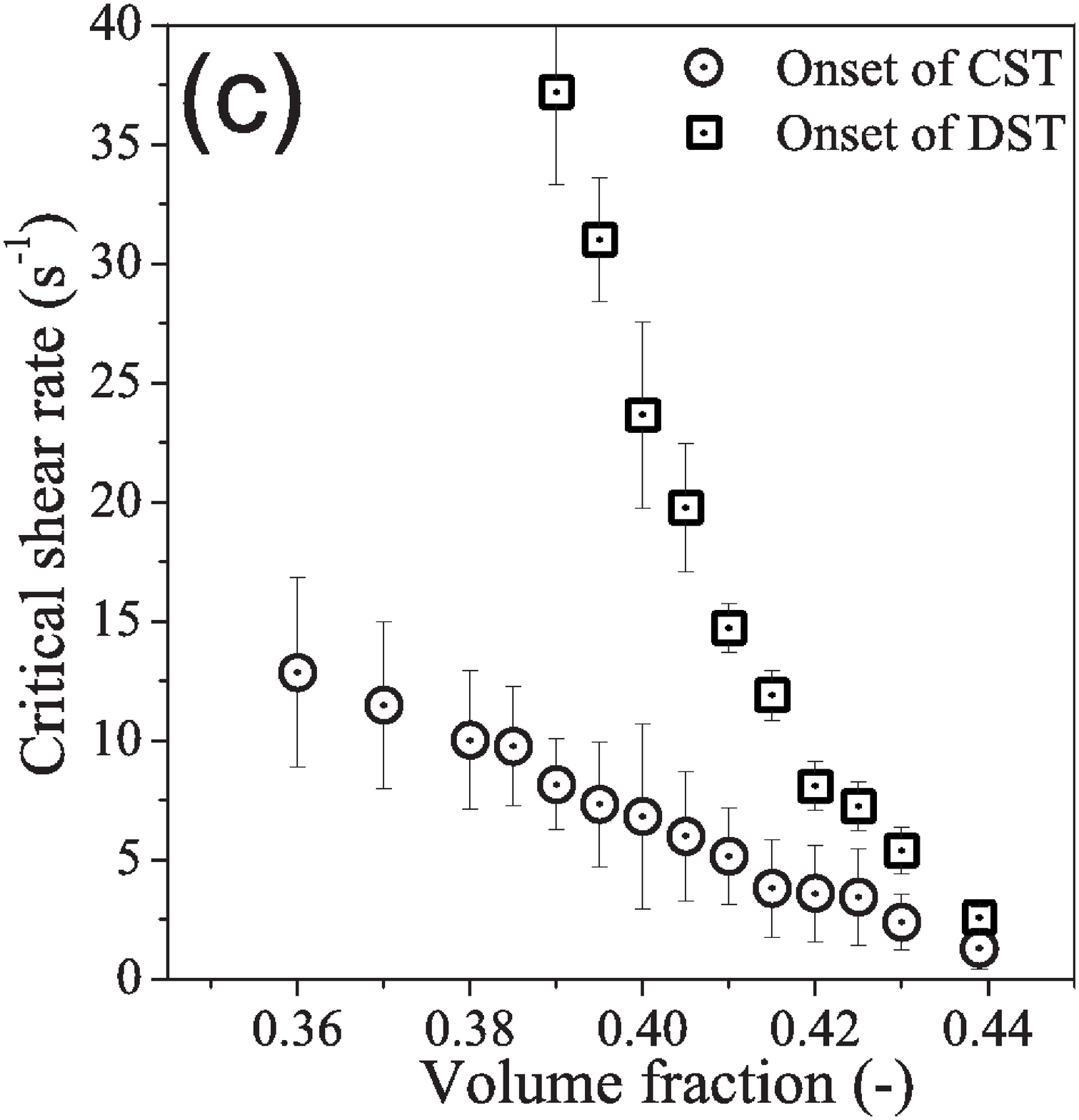}\hfil
\includegraphics*[width=0.23\textwidth]{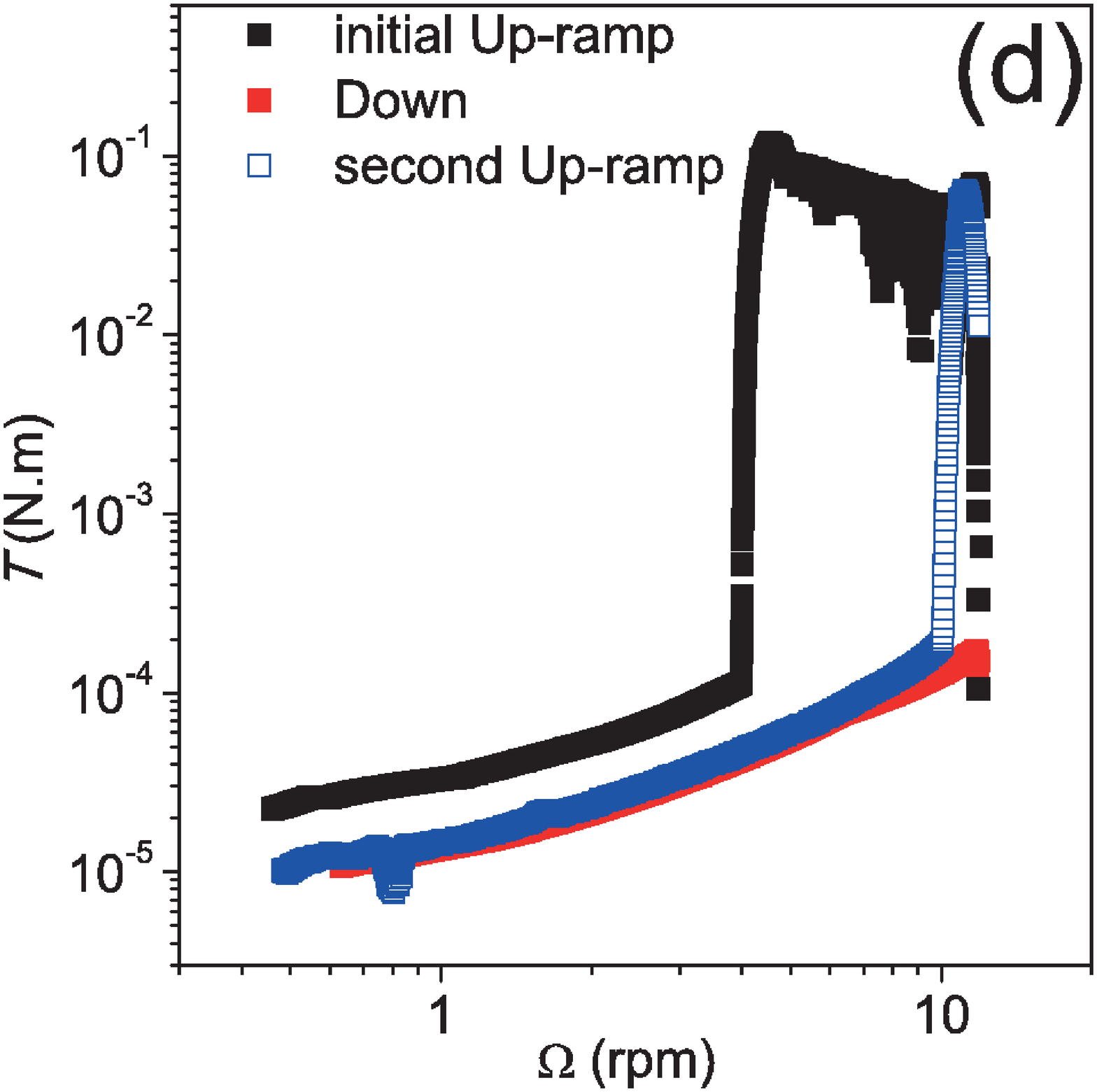}
\caption{
\label{fig:ramp} (Color online). Macroscopic rheometry data. (a) Torque $T$ vs rotation rate $\Omega$ during a logarithmic ramp at various packing fractions $\phi$. (b) Apparent viscosity vs apparent shear rate extracted from (a). (c) Critical shear rates $\dot\gamma_{\rm CST}$ and $\dot\gamma_{\rm DST}$ vs $\phi$. (d) Reversibility test at $\phi=43.9\%$: succession of up, down, and up ramps.
}
\end{figure}

Let us start with macroscopic rheometry measurements using a {\it Malvern Kinexus Pro\/} stress-controlled Couette rheometer. The inner and outer cylinders radii are: $R_i=12.5$mm and $R_o=18$mm; the inner cylinder height is $h=37.5$mm; the gap is large enough to avoid confinement effects~\cite{FallHuangBertrandOvarlezBonn2008,FallBertrandOvarlezBonn2012,BrownJaeger2012}. Both cylinders are roughened to avoid wall slip.
We report on Fig.~\ref{fig:ramp}-(a) the torque $T$ measured on the inner cylinder during a 3 min logarithmic ramp of $\Omega$ (the inner cylinder rotation rate) from $5.10^{-2}\,{\rm rpm}$ to $2.10^3\,{\rm rpm}$. Near the inner cylinder, the local stress is
\begin{equation}\label{eq:T}
\tau(R_i,\Omega)=T(\Omega)/(2\,\pi R_i^2h)
\end{equation}
the local strain rate is a priori unknown, but is classicaly estimated as:
\begin{equation}\label{eq:gdot}
\dot\gamma(R_i,\Omega)\simeq2\Omega\,R_o^2/(R_o^2-R_i^2)
\end{equation}
Combining Eq.~(\ref{eq:T}) and~(\ref{eq:gdot}) yields the apparent viscosity $\eta\equiv\tau/\dot\gamma$ vs $\dot\gamma$ relation reported on Fig.~\ref{fig:ramp}-(b). Note that at low $\eta$ macroscopic inertial effects may arise (Taylor-Couette instabilities), which limit the accessible $\dot\gamma$ range at low $\phi$'s.

At low $\dot\gamma$, the response is clearly shear-thinning. It crosses over to CST around a characteristic strain rate $\dot\gamma_{\rm CST}$ defined at that where $\eta$ reaches a minimum. Viscosity jumps at some higher strain rate $\dot\gamma_{\rm DST}$ which characterizes the onset of DST. 

The classical estimate of $\dot\gamma(R_i)$ as Eq.~(\ref{eq:gdot}), which is based on the Newtonian solution, remains unsatisfactory. An exact expression exists so long as the material is homogeneous in the studied torque range:
\begin{equation}\label{eq:sum}
\dot\gamma(R_i;T_0)=2\sum_{n=0}^\infty\,\left.\left(T\frac{\d\Omega}{\d T}\right)\right|_{T=T_0\left(\frac{R_i}{R_o}\right)^{2n}}
\end{equation}
It cannot be used around the DST transition, however, because the apparent singularity of ${\d T}/{\d\Omega}$ cannot be resolved experimentally. Nevertheless we have checked that using the first two terms of~(\ref{eq:sum}) gives the same qualitative behaviour as in Fig.~\ref{fig:ramp}-(b) up to DST. We will show below that the flow does remain homogeneous below DST, which allows us to use~(\ref{eq:sum}) to accurately compute the $\dot\gamma_{\rm CST}$ and $\dot\gamma_{\rm DST}$ values that are reported on Fig.~\ref{fig:ramp}-(c). Interestingly, (i) at any volume fraction CST is always observed before DST; (ii) both $\dot\gamma_{\rm CST}$ and $\dot\gamma_{\rm DST}$ seem to vanish around the same volume fraction $\phi_c$ which we estimate to be $\approx45\%\ll\phi_{\rm RCP}$ by linear extrapolation.

Now we turn to velocity-controlled MRI rheometry. Our Couette cell has inner and outer radii $R_i=3$cm and $R_o=5$cm (resp.) and inner cylinder height $h=11$cm. Both cylinders are roughened to avoid slip, which we checked from velocity profiles. All experiments discussed below are performed by preparing a homogeneous material with mean volume fraction $\phi_0=43.9\%$ (experiments at 40\%, 41\% and 42.5\% show similar features). Note that our MRI cell dimensions differ from those of the cell used to obtain the macroscopic data of Fig~\ref{fig:ramp}. Hence, to capture DST with this new cell, the range of rotation velocities must be adapted: at $\phi_0=43.9\%$, we use $\Omega$ values between $5\,{\rm rpm}$ and $100\,{\rm rpm}$. In each MRI experiment, $\Omega$ is held fixed until steady state is reached.

Our MRI~\cite{RodtsBertrandJarnyPoullainMoucheront2004,OvarlezBertrandRodts2006} provides the stationary packing fraction $\phi(r,\Omega)$ and azimuthal velocity $v(r,\Omega)$ at any radial position $r$. From the latter we extract the local strain rate $\dot\gamma(r,\Omega)=v/r-{\partial v}/{\partial r}$. We do not have access to torque measurement. But, since the local stress is $\tau(r,\Omega)=\tau(R_i,\Omega)R_i^2/r^2$ in the Couette geometry, we can estimate the local viscosity profile as:
\begin{equation}\label{eq:viscosity}
\eta(r,\Omega) = \frac{\tau(R_i,\Omega)}{\dot\gamma(r,\Omega)}\,\frac{R_i^2}{r^2}
\end{equation}
up to the unknow prefactor $\tau(R_i,\Omega)$.

\begin{figure}[t]
\includegraphics*[width=0.35\textwidth]{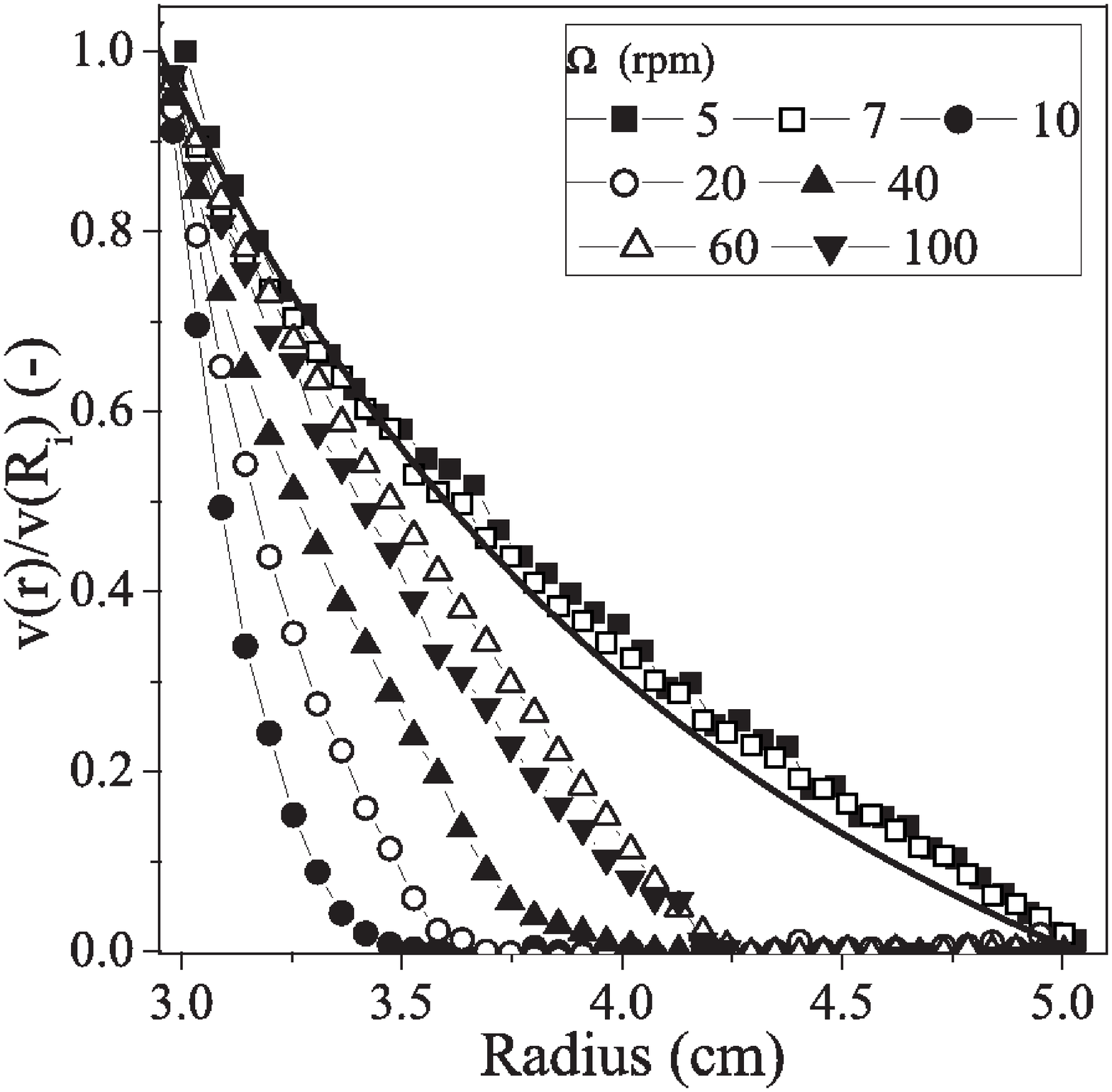}\hfil
\includegraphics*[width=0.35\textwidth]{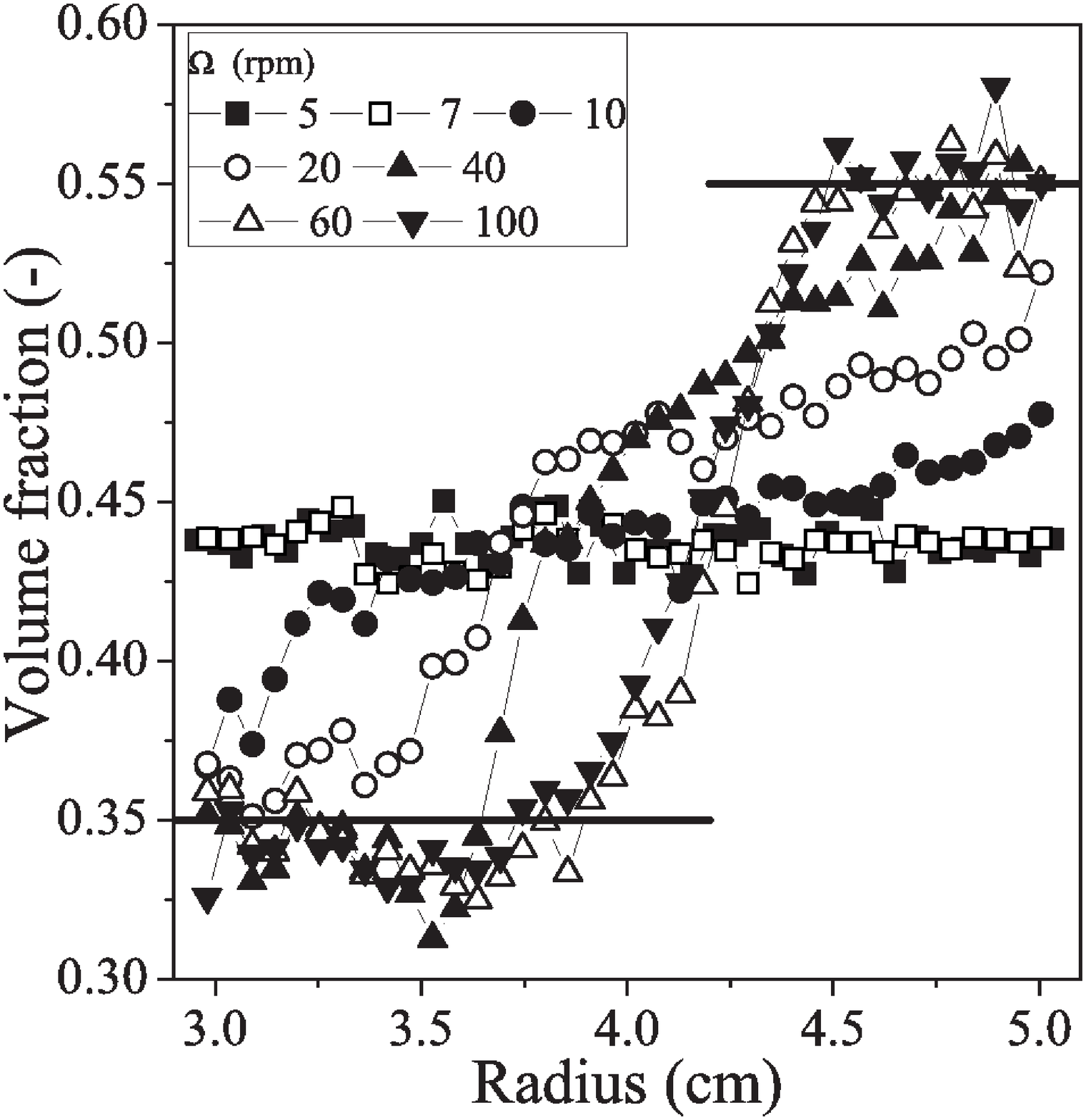}
\caption{
\label{fig:profiles}
Steady MRI data for a $\phi_0=43.9\%$ cornstarch suspension and different rotational velocities $\Omega$. (a) velocity profiles. (b) density profiles; solid lines indicate $\phi_{\rm RLP}$ and $\phi_{\rm RCP}$.
}
\end{figure}

In Fig~\ref{fig:profiles}, we plot the steady velocity and concentration profiles thus measured for a few $\Omega$s. Velocity is normalized by its value at the inner cylinder.

\begin{figure}[t]
\includegraphics*[width=0.23\textwidth]{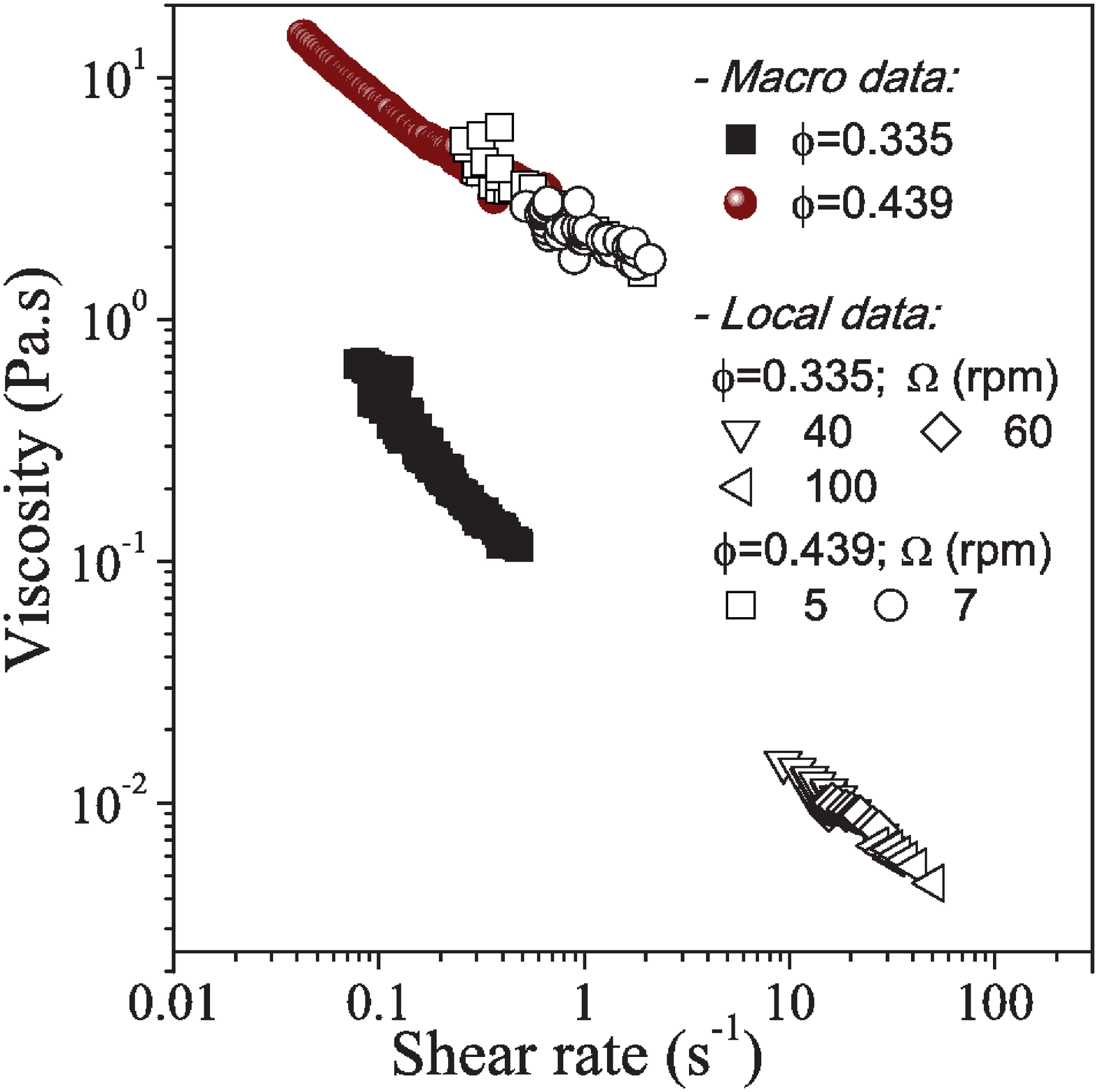}\hfil
\includegraphics*[width=0.23\textwidth]{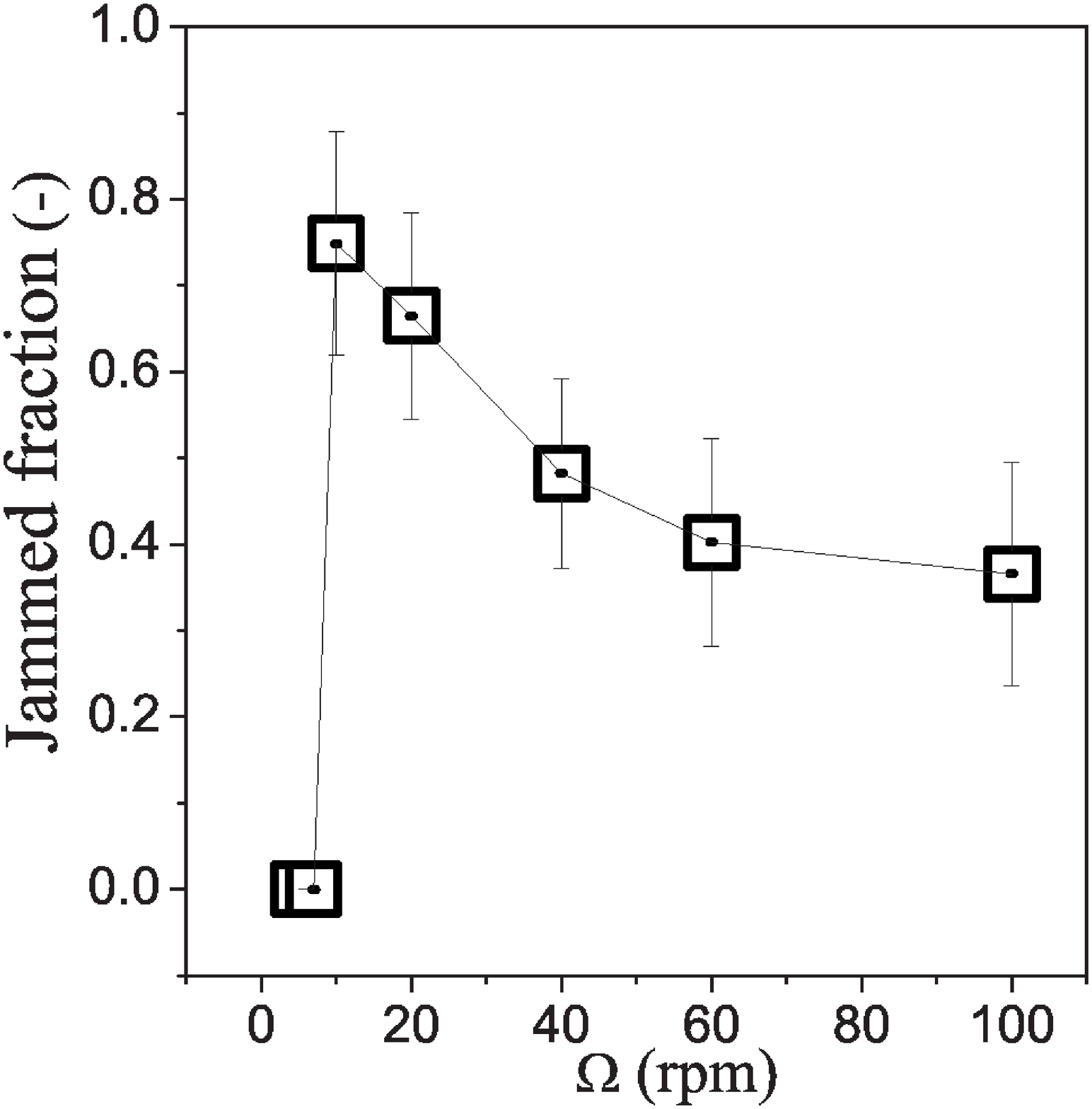}
\caption{
\label{fig:analysis}
(a) Comparison of local rheometry data obtained from MRI measurements (open symbols) and near the inner cylinder in macroscopic rheometry (filled symbols) in homogeneous conditions. Upper data: $\phi=43.9\%$; lower data: $\phi\approx33.5\%$.
(b) From the velocity profiles [Fig.~\ref{fig:profiles}-(a)]: fraction of the gap which is jammed vs $\Omega$.
}
\end{figure}

For small $\Omega$'s (5 and 7~rpm), we find that the density profile is homogeneous while the flow extends throughout the gap. It is known that non-Brownian suspensions may slowly become inhomogeneous due to shear-induced migration and sedimentation~\cite{OvarlezBertrandRodts2006,FallBertrandOvarlezBonn2009}. We checked that this does not occur before strains larger than a few thousands, which is much larger than the strain range over which we collect data. Since density is uniform, we can access the local rheology $\eta(\dot\gamma,\phi_0)$ as follows. For each $\Omega$, matching the single parameter $\tau(R_i,\Omega)$ in Eq.~(\ref{eq:viscosity}) provides the complete $\eta(r,\Omega)$ profile, which can thus be plotted vs $\dot\gamma(r,\Omega)$. These values are compared on Fig.~\ref{fig:analysis}-(a) with the local rheology data obtained from macroscopic rheometry using Eqs.~(\ref{eq:T}) and~(\ref{eq:sum}) to evaluate the local stress and strain rate near the inner cylinder. Clearly, in the shear-thinning regime, the stationary local response within the gap matches our macroscopic measurements

This local homogeneous response is observed so long as the maximal local strain rate, which is reached at the inner cylinder, lies below the $\dot\gamma_{\rm DST}(\phi_0)$ value identified in macroscopic rheometry [Fig.~\ref{fig:ramp}-(c)]. Homogeneity and locality then enable us to estimate the critical rotation rate at which DST is expected as: $\Omega_{\rm DST}(\phi_0)\simeq8$rpm.

A sudden transition occurs as soon as $\Omega$ crosses $\Omega_{\rm DST}(\phi_0)$. As shown on Fig.~\ref{fig:profiles}-(a), the flow then abruptly stops in a large region. Namely, the velocity profile jumps from one of the rightmost curves, corresponding to homogeneous flows, to the leftmost one, i.e. the most strongly localized flow. Note that measuring a single velocity profile requires accumulating MRI data over $\sim30$s, which corresponds here to a strain of order 50.  Upon crossing $\Omega_{\rm DST}(\phi_0)$, the first measurable velocity profile is already localized. The flow subsequently remains steady over thousands units of strains. DST is thus clearly concomitant with shear localization.

As $\Omega$ increases further, the velocity profiles progressively extend to the right (i.e. towards the outer cylinder). In all cases, the system remains separated into a flowing layer near the inner cylinder and a jammed region near the outer one. The fraction of the gap that is jammed is reported on Fig.~\ref{fig:analysis}-(b): it jumps at $\Omega_{\rm DST}(\phi_0)$ and then slowly decreases.

Comparing these velocity profiles [Fig.~\ref{fig:profiles}-(a)] with local density data [Fig.~\ref{fig:profiles}-(b)], we find that, quite strikingly, the flow localization at $\Omega=10\text{rpm}\gtrsim\Omega_{\rm DST}(\phi_0)$ is associated with the sudden emergence of density inhomogeneities. Namely, the volume fraction decreases in the flowing layer, while it increases in the jammed region, as required by the conservation of particle number. 
As $\Omega$ increases beyond $\Omega_{\rm DST}(\phi_0)$, the progressive extension of the flowing layer is accompanied by a broadening of the low-density region. At high strain rates, the density saturates, in the flowing layer, at a packing fraction $\phi_{\rm min}\simeq33\%\lesssim35\%\simeq\phi_{\rm RLP}$ and, in the jammed region, at $\phi\sim\phi_{\rm RCP}$. It is noteworthy that the density profile can achieve multiple forms depending on shear history~\footnote{This is in stark contrast with the system of~\cite{FallLemaitreBertrandBonnOvarlez2010}, where density is stationary after DST.}.

Let us emphasize that the change of density created by the DST transition is irreversible. Indeed, once a stationary profile $\phi(r,\Omega_1)$ has been produced by ramping $\Omega$ up to some arbitrary $\Omega_1>\Omega_{\rm DST}$, we found that the density profile remains the same under any subsequent \emph{lowering} of $\Omega$. This irreversibility shows up in our macroscopic rheometry setup (the small Couette cell) as illustrated on Fig.~\ref{fig:ramp}-(d) where we plot the torque $T$ vs $\Omega$ during: (i) an initial up-ramp that drives the system through the DST transition; followed by (ii) a down-ramp. The torques measured during the up- and down-ramps clearly lie on different branches. However, if we subsequently (iii) re-increase $\Omega$, torque $T(\Omega)$ tracks the data previously obtained on the down-ramp. Hence we reason that on the down (ii) and up (iii) ramps, the system explores quasi-reversibly a family of states associated with some inhomogeneous density profile set by $\Omega_1$, the maximum value of the rotation rate achieved during shear history.

We now probe the local rheology within the shear-thickened flow. Three of our datasets, for $\Omega=40$, 60, and 100rpm, present a finite-width region of roughly uniform packing fraction $\phi_{\rm min}\simeq33\%$. Using only the data collected in these regions, we can hence implement the same approach as in the fully homogeneous case, to access the local rheology $\eta(\dot\gamma,\phi)$ at $\phi=\phi_{\rm min}$. For each $\Omega$, we match the prefactor $\tau(R_i,\Omega)$ and plot in Fig.~\ref{fig:analysis}-(a) the resulting $\eta(r,\Omega)$ vs $\dot\gamma(r,\Omega)$ data points. Clearly, the local $\eta(\dot\gamma,\phi_{\rm min})$ rheology is shear-thinning, and extrapolates the $\phi=33.5\%$ rheometry measurements. 

These data demonstrate that macroscopic DST cannot be straightforwardly interpreted as capturing a comparable change of the local response at fixed volume fraction. Instead, it results from the separation of the system into a low-density-flowing and a high-density-jammed regions. Somewhat paradoxically, the local response was found above to be shear-thinning in the flowing layer of the macroscopic shear-thickened state. 

\begin{figure}[t]
\includegraphics*[width=0.45\textwidth,height=0.4\textwidth]{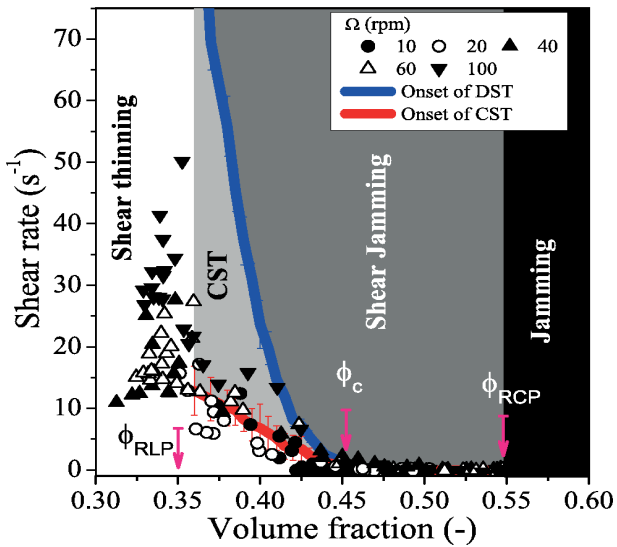}
\caption{
\label{fig:phases}
Phase diagram: scatter plot of the local $(\dot\gamma(r,\Omega), \phi(r,\Omega))$ values compared with the $\dot\gamma_{\rm CST}(\phi)$ (red) and $\dot\gamma_{\rm DST}(\phi)$ (blue) lines obtained from macroscopic rheometry [Fig.~\ref{fig:ramp}-(c)].
}
\end{figure}

In the flowing layer, the density plateau only develops around $\phi_{\rm min}$. It is hence at this packing fraction only that we can extract the local rheology $\eta(\dot\gamma,\phi)$ from MRI data. To further qualify the local response, we now construct a phase diagram for all the local flow and jammed states. To this end, we collect the operating points $(\dot\gamma(r,\Omega),\phi(r,\Omega))$ obtained with the MRI for all $\Omega$'s and all $r$'s and report them on Fig.~\ref{fig:phases}, where they are compared with the curves $\dot\gamma_{\rm CST}(\phi)$ and $\dot\gamma_{\rm DST}(\phi)$ that delimit the flow regimes previously identified using macroscopic rheometry. Our MRI data include points obtained in the flowing layer as well as in jammed regions: in the scatter plot, they are respectively found on the left (low $\phi$, with non-vanishing $\dot\gamma$) and right (high $\phi$, $\dot\gamma\simeq0$) hand sides.

Local flow point fall in either the shear-thinning or CST domains, which supports that both types of local rheologies are found. Moreover, they all lie below the $\dot\gamma_{\rm DST}(\phi)$ line, which thus appears to be an intrinsic material limit beyond which no steady flow can be sustained. No flow is found in particular at packing fractions above the value $\phi_c\simeq45\%$ where $\dot\gamma_{\rm CST}(\phi)$ and $\dot\gamma_{\rm DST}(\phi)$ vanish. As $\phi_c$ is quite distinct from the jamming point $\phi_{\rm RCP}\simeq55\%$, a continuum of jammed states is accessible beween $\phi_c$ and $\phi_{\rm RCP}$.

In the literature, experimental observations of macroscopic DST are usually analyzed while assuming that the flow remains homogeneous: the DST stress jump hence is interpreted as picking up a secondary, high-viscosity, branch of the steady local response. As we have shown, however, homogeneity does not hold. The stress jump is a property of the global response, and is associated with a global reorganization of the flow, while the local response always lies within the shear-thinning or CST regimes.
Of course, this is shown here for cornstarch, but we note that today, there is not in the literature, a single experimental data point that can be trusted to pick up the presumed secondary branch of s-shaped rheologies, since there is no other local observation of the flow response. 


Moreover, the DST scenario that we are unveiling here is rather unexpected in view of recent models based on s-shape flow curves~\cite{WyartCates2014}. Indeed, at the onset of macroscopic DST, the part of the flow that reaches locally a limit of stability, i.e. the $\dot\gamma_{\rm DST}$ line, is the region near the inner cylinder, where the local shear rate is maximal. From s-shape models, in their current formulation, we would expect this locally unstable region to jump to a high-viscosity state. In stark contrast, we find that when DST occurs, this region jumps toward a low-density, low-viscosity shear-thinning state. Meanwhile jamming occurs in the region near the outer cylinder which, at the onset of DST, was well within its stability limit.


It finally appears that DST is determined by two major factors: (i) the existence of a limit $\dot\gamma_{\rm DST}(\phi)$ beyond which no steady flow can be accessed; (ii) the fact that as $\phi_c\ll\phi_{\rm RCP}$, a continuum of jammed states exists. Indeed, due to (i) the inner cylinder region eventually reaches the $\dot\gamma_{\rm DST}(\phi)$ shear-jamming limit, that it can only escape by decreasing its local volume fraction; it thus drives migration throughout the gap, because (ii) the growing jammed layer can store material significantly above $\phi_c$. Both~(i) and~(ii) are direct consequences of a unique physical phenomenon, shear jamming.

It is crucial here that the $\dot\gamma_{\rm DST}(\phi)$ line is not vertical, hence can be crossed by increasing $\dot\gamma$ at given $\phi$. It is so because the asymptotic flow regimes (high and low $\dot\gamma$, resp.) have two different limit volume fractions ($\simeq\phi_{RLP}$ and $\phi_c$, resp.). This contrasts sharply with~\cite{FallLemaitreBertrandBonnOvarlez2010} where jamming is rate-independent and DST purely dynamical.
It remains open to identify the microscopic mechanisms that account for the splitting of the jamming limit observed here.
Several may be envisionned, such as the increased mobilization of frictional contacts with increasing shear rate~\cite{SetoMariMorrisDenn2013}, or the angularity of particles, which is significant for cornstarch, or finally particle inertia, which might be causing the CST~\cite{FallLemaitreBertrandBonnOvarlez2010} that always preceeds DST.





%

\end{document}